# Symptoms of depersonalisation/derealisation disorder as measured by brain electrical activity: A systematic review

Abbas Salami[a*], Javier Andreu-Perez[b], Helge Gillmeister[c]

a. School of Computer Science and Electronic Engineering, University of Essex, Wivenhoe Park, Colchester, CO4 3SQ, UK, a.salami@essex.ac.uk
b. School of Computer Science and Electronic Engineering, University of Essex, Wivenhoe Park, Colchester, CO4 3SQ, UK, javier.andreu@essex.ac.uk
c. Department of Psychology and Centre for Brain Science, University of Essex, Wivenhoe Park, Colchester, CO4 3SQ, UK, helge@essex.ac.uk

## Abstract

Depersonalisation/derealisation disorder (DPD) refers to frequent and persistent detachment from bodily self and disengagement from the outside world. As a dissociative disorder, DPD affects 1-2% of the population, but takes 7- 12 years on average to be accurately diagnosed. In this systematic review, we comprehensively describe research targeting the neural correlates of core DPD symptoms, covering publications between 1992 and 2020 that have used electrophysiological techniques. The aim was to investigate the diagnostic potential of these relatively inexpensive and convenient neuroimaging tools. We review the EEG power spectrum, components of the event-related potential (ERP), as well as vestibular and heartbeat evoked potentials as likely electrophysiological biomarkers to study DPD symptoms. We argue that acute anxiety- or trauma-related impairments in the integration of interoceptive and exteroceptive signals play a key role in the formation of DPD symptoms, and that future research needs analysis methods that can take this integration into account. We suggest tools for prospective studies of electrophysiological DPD biomarkers, which are urgently needed to fully develop their diagnostic potential.

[*] Corresponding author.
E-mail address: a.salami@essex.ac.uk

# Introduction

Depersonalisation/derealisation refers to a state of mind in which a person feels detached and disconnected from their bodies and own senses as well as from their surroundings (Phillips and Sierra, 2003). This condition can be accompanied by a sense of unreality about the outside world and by out-of-body experiences, and is believed to be a defensive mechanism of the brain in order to protect the organism in acute anxiety or traumatic situations (Hunter et al., 2003; Shilony and Grossman, 1993; Sierra and Berrios, 1998; Stein and Simeon, 2009). For instance, temporary occurrences of depersonalisation have been reported by almost 50% of college students (Dixon, 1963). Fatigue (Tibubos et al., 2018), sleep deprivation (van Heugten–van der Kloet et al., 2015), or travelling to unfamiliar places can also be the cause of transient depersonalisation/derealisation (Kaplan et al., 1998). However, in cases where the symptoms are chronic, it is considered a type of dissociative disorder (depersonalisation/derealisation disorder (DPD); Diagnostic and Statistical Manual of Mental Disorders, 5th Edition (DSM-5) (American Psychiatric Association, 2013)).

Although the exact cause of DPD is not yet known, traumatic experiences and childhood anxiety are thought to be common triggers (Lee et al., 2012; Simeon et al., 2001b). It can also be provoked by intense stress, depression, panic attacks, and ingestion of psychoactive substances (Mathew et al., 1999; Medford et al., 2003; Simeon, 2004). Patients experience persistent and frequent feelings of disembodiment or detachment from their physical self as well as emotional numbness that may extend beyond the present moment to include memories and imagination. Since derealisation is an inseparable aspect of DPD in most cases, the symptoms may also include detachment from surroundings, as if the world around the patient is unreal, or a weakened ability to respond to emotional circumstances, although the capacity for emotional expression and reality testing remains intact (Hollander et al., 1992; Sierra and David, 2011). DPD can be accompanied by anxiety, depression or schizophrenia (Stein et al., 1997), as well as difficulties in concentration and memory retrieval (Lambert et al., 2001a), which can profoundly affect the quality of life for patients and interfere with their daily activities and social relationships. Identifying DPD, as well as its risks and neuroprotective factors, at early stages should thus be a critical endeavour for clinical practice and research.

DPD has a prevalence of about 1-2% of the population (Hunter et al., 2004; Lee et al., 2012; Michal et al., 2007), which is comparable to that of schizophrenia and obsessive-compulsive disorder, with an equal gender ratio and an average onset age in early adulthood (Baker et al., 2003; Michal et al., 2016; Simeon, 2004; Simeon et al., 2003). Nevertheless, DPD is one of the most prevalent but under-diagnosed psychiatric disorders (Hunter et al., 2017; Michal et al., 2010). Generally, there are no medical laboratory tests for diagnosis of dissociative disorders, and since patients find it difficult to describe the symptoms of depersonalisation and derealisation, it currently takes an average of seven to 12 years to correctly diagnose DPD (Hunter et al., 2017). Diagnosis is hampered by a lack of awareness of DPD among medical practitioners (Medford et al., 2005) and its symptomatology overlap with medical conditions such as epilepsy and migraine (Devinsky et al., 1989; Lambert et al., 2002) and psychiatric conditions such as depression and post-traumatic stress disorder (Armour et al., 2014). Therefore, delineating the neurophysiological correlates of DPD may be of great importance for an early diagnosis of DPD as discriminate it from its transient form and from other conditions.

In this review, we provide an overview of the structural and functional neurophysiology in DPD, with a particular focus on studies aiming to characterise the cardinal symptoms of the disorder such as feelings of disembodiment and emotional numbing by measuring the electrical activity of the brain. Electrophysiological neuroimaging techniques are of great interest because of their ease of application and cost-effectiveness for clinical practice. Therefore, we intend to identify and introduce electrophysiological biomarkers associated with DPD symptoms, which may have the potential to help with early recognition of this under-diagnosed psychiatric condition. Our paper demonstrates both the urgent need to replicate promising findings on a larger scale and the potential for further electrophysiological pattern analysis to characterise DPD.

## Functional Nervous System Studies of DPD

Several methods have been used to measure the functional characteristics of DPD symptoms within the autonomic nervous system (galvanic skin response, heart rate) and the central nervous system (EEG, ERPs, HEPs, fMRI, PET). The Galvanic Skin Response (GSR) tracks changes in the conductivity of human skin due to sweating, reflective of the arousal related to the intensity of emotional states. This property has made GSR a useful approach to investigate emotional dysregulations in DPD. Studies using the GSR have elucidated differential autonomic nervous system responses to salient stimuli in DPD, and the impairment is mostly in responding to unpleasant and threatening emotional stimuli rather than pleasant ones (Michal et al., 2013; Monde et al., 2013; Sierra et al., 2006). For instance, research by Dewe et al. (Dewe et al., 2016) observed suppressed GSRs during the reception of body-threat-like stimuli (blood donation) in DPD patients. Giesbrecht et al. (Giesbrecht et al., 2010) recorded a different temporal pattern of GSRs in DPD patients compared to healthy individuals while watching an anxiety-inducing movie scene. The response of the patients was characterized by an early peak and a flattened pattern afterwards (even after clip offset), while the control group exhibited a more gradual incremental pattern during the clip followed by a decreasing pattern after clip offset. Peak latency was inversely associated with the severity of DPD symptoms, and patients also showed higher baseline skin responses compared to the control group. Jay et al. (Jay et al., 2014) measured DPD patients' galvanic skin responses and used repetitive transcranial magnetic stimulation (rTMS) to confirm the causal role of ventrolateral prefrontal cortex in these atypical patterns of physiological arousal.

Several neuroimaging studies in the literature have targeted central neural patterns and possible abnormal activities in DPD with functional magnetic resonance imaging (fMRI) (Phillips and Sierra, 2003; Röder et al., 2007), positron emission tomography (PET) (Simeon et al., 2000), and electroencephalography (EEG). These studies predominantly compare the neural substrates of DPD patients with control subjects and have mainly focused on two core aspects of DPD, emotional numbing and disembodiment (Simeon et al., 2008). For instance, various fMRI studies (Lemche et al., 2008; Lemche et al., 2007; Mancini-Marïe et al., 2006) have investigated the neural responses of DPD patients to emotional versus neutral stimuli (Medford et al., 2006; Phillips et al., 2001). Results illustrate that emotional numbing (the attenuation of emotional experiences as a result of inhibitory processes) in DPD is associated with reduced activity in brain areas responsible for emotional processing, particularly the insula and limbic regions including hypothalamus and amygdala (Medford et al., 2016). Lemche et al. (Lemche et al., 2007) showed an inverse relationship between activity in hypothalamus and amygdala and the intensity of positive and negative emotional stimuli in a group of DPD patients compared with controls. fMRI studies also showed increased activation of right ventrolateral prefrontal cortex in DPD patients exhibiting emotional numbness in response to aversive stimuli (Medford et al., 2006; Phillips et al., 2001). Similarly, increased activation of dorsal prefrontal cortex, which plays a role in emotional suppression (Etkin et al., 2006), was found during the processing of both positive and negative emotional facial expressions in DPD patients (Lemche et al., 2007). Dorsal prefrontal activation was inversely related to skin conductance levels. This suggests that prefrontal regions actively suppress the operation of emotional cortical and limbic regions. In line with this proposal, fifteen minutes of 1Hz inhibitory rTMS to the right ventrolateral prefrontal cortex was found to result in an increase of skin conductance capacity, which reflects the capacity of autonomic responses to emotional stimuli (Jay et al., 2014).

Investigations of disembodiment, another critical (but highly distinctive) characteristic of DPD, can be found less frequently in the DPD literature. Nevertheless, Paul et al. (Paul et al., 2019) recently conducted a comprehensive fMRI study on the functional connectivities between brain areas that might be associated with depersonalisation/derealisation symptoms (such as extrastriate body area, hippocampus, medial prefrontal cortex, and posterior and anterior insular cortex) in patients with major depressive disorder. Results revealed that decreased functional connectivity between extrastriate body area (which plays a role in the perception of body parts) and default mode network (which is associated with the processing of self-relevant, autobiographical information) is related to depersonalisation/derealisation symptoms in major depressive disorder. Altered functional connectivity of the default mode network and particular areas in the primary visual network has also been discovered by Derome et al. (Derome et al., 2018) in transient depersonalisation.

Among the neuroimaging techniques for the analysis of brain activity, EEG holds great promise as a diagnostic tool because of its non-invasive nature, low costs and simple setup. It provides information about the ongoing neural processes in the cerebral cortex with high temporal resolution. This paper aims to provide a review of studies on DPD using electrophysiological signals to detect and introduce electrophysiological biomarkers associated with DPD symptoms. It also addresses some recent developments in the theories of self-consciousness that can potentially help to explain the unique symptomatology of DPD.

## Review Methodology

The papers reported in this article are exclusively based on electrophysiological approaches, and comprises of all the papers that have tried to explain symptoms of DPD using scalp electrophysiological signals (for review of studies on DPD using other behavioural or neuroimaging techniques see (Hunter et al., 2004; Sierra and David, 2011)). We categorized our search based on the four major and distinct symptoms of DPD derived from factor analysis on the Cambridge Depersonalisation Scale (CDS) (Sierra et al., 2005; Sierra and Berrios, 2000), which is the most well-known benchmark questionnaire for quantification of DPD (Talbert, 2010). In addition, we describe studies related to 'other symptoms' of DPD, such as those related to perception, attention and working memory (see **Error! Reference source not found.**). We also explicitly distinguished between papers based on whether they investigated transient or chronic DPD, with transient DPD defined as episodes of depersonalisation in healthy individuals and those with a primary diagnosis of another illness. Since the essence of depersonalisation is a self-protective mechanism, its momentary symptoms can emerge in many healthy individuals during their lifespan as the brain's response to an overwhelmingly stressful or traumatic situation in order to reduce its repercussions by creating a sense of physical and emotional numbness. Studies have reported the prevalence of transient depersonalisation in the range of 34 to 70% in the non-clinical population (Aderibigbe et al., 2001; Hunter et al., 2017). Besides, although a recent comprehensive study has confirmed DPD as a distinct disorder (Sierra et al., 2012), transient depersonalisation and derealisation are also common symptoms in several major psychiatric illnesses such as anxiety (Baker et al., 2003), panic attacks (Harper and Roth, 1962), burnout syndrome (Maslach and Jackson, 1981), and post-traumatic stress disorder (Lanius et al., 2012). For instance, episodic depersonalisation attacks along with panic attacks have been jointly observed in several cases (Davison, 1964; Dietl et al., 2005) and this symptomatic pattern has been considered a distinct disorder, the "phobic anxiety-depersonalisation syndrome" (Roth, 1960). The dissociative subtype of post-traumatic stress disorder has been distinguished from its nondissociative subtype symptomatically and by distinct patterns of central nervous system activities (Lanius et al., 2010). In the dissociative subtype, patients show episodic depersonalisation symptoms, which correlate negatively with the activation of amygdala and right anterior insula (Hopper et al., 2007).

## Search Keywords and Information Sources

Several combinations of keywords were used in this systematic search. The search keywords consisted of one word from Set 1 and one word from Set 2 as follows (notice that unlike Scopus, distinct results will be derived for British and American spelling of the keywords in Google Scholar):

Set 1: "Depersonalization", "Depersonalisation", "Derealization", "Derealisation"

Set 2: "Electrophysiological", "EEG", "MEG", "Biomarker", "Interoception", "Exteroception"

The search was conducted in two major electronic databases, Google Scholar and Scopus. The search was not restricted to any year ranges and selected papers covered publications from 1992 to February 2020. References of all the relevant papers were also scanned to find further potential studies in the field.

### Articles Overview

Papers selected (based on their title) from Google Scholar and Scopus created an initial database of 104 studies after removing duplicate papers. The papers were manually filtered and the number reduced to 70 based on their abstracts. Papers concerned with (neuroimaging) techniques other than electrophysiology were excluded from the review. A final assessment of papers was conducted based on an evaluation of the full text. The final number of selected papers was 10, and these are presented as an overview in **Error! Reference source not found.**, and described in detail in the next section.

## Electrophysiological Studies of DPD

Earlier studies sought to investigate the oscillatory signatures associated with the experience of depersonalisation. Locatelli et al. (Locatelli et al., 1993) examined EEG patterns in depersonalisation states. They aimed to observe the probable dysregulation in the temporolimbic regions of the brain in healthy subjects and panic disorder patients with and without depersonalisation/derealisation using an odour discrimination task. The power of the EEG signals in eight separate frequency bands covering 1 to 30Hz was analysed at six temporal electrodes. The results revealed bilateral abnormalities in EEG of the temporal lobe in patients with depersonalisation as compared with panic disorder patients without depersonalisation or derealisation. Patients with depersonalisation/derealisation showed increased power in the delta band and a bilateral lack of responsiveness in the upper alpha band during odour stimulation. Hollander et al. (Hollander et al., 1992) reported EEG power changes in depersonalisation in a different frequency band. They investigated the neurophysiological basis of depersonalisation in a 23-year-old man, who had reported depersonalisation and derealisation symptoms after suffering from severe anxiety for a period. Although the long-term resting-state EEG was reported as normal, brain electrical activity mapping revealed frontal alpha overactivation and increased temporal theta activity in the left hemisphere. They also reported enhanced N200 components of visual and auditory ERPs (~200ms post-stimulus) in the left temporal areas. DPD was frequently found to be associated with abnormal theta activity over temporal regions. In an alcohol-induced depersonalisation state, Raimo et al. (Raimo et al., 1999) observed a significantly higher relative power in theta band compared with asymptomatic episodes. The presence of abnormal theta activities in depersonalisation state was also confirmed in (Hayashi et al., 2010) and associated with a large effect size. Hayashi et al. (Hayashi et al., 2010) showed that depersonalisation symptoms in panic disorder patients could induce abnormalities in the EEG pattern of patients, characterised by repeated slow wave (in the range of theta) bursts.

## Disembodiment feelings (desomatisation)

One of the cardinal symptoms of depersonalisation is disembodiment – feeling detached or estranged from one's own body parts or whole body, when looking at them directly or in a mirror. DPD patients also often complain about a lack of agency - feeling as if their speech or movements are robotic and not their own. In other words, the disorder is characterized by frequent and persistent experiences of a loss of the physical sense of self (the feeling of oneself as the bodily subject of one's experiences). Adler et al. (Adler et al., 2016) tried to explain disembodiment in DPD patients using a somatosensory resonance paradigm, which taps into the human mirror neuron system (Iacoboni et al., 1999; Molenberghs et al., 2012). The mirror neuron system, which resides in the premotor cortex and inferior parietal cortex as well as associated regions of sensory, motor and emotional processing (Molenberghs et al., 2009), is active both when we perform an action (action execution) and when we observe someone else engaging in a similar action (action observation) (Morales et al., 2019) and is thought to encode the functional goal of an action (Le Bel et al., 2009). The mirror neuron system plays a crucial role in the adult representation of the bodily self (Hu et al., 2016; Molnar-Szakacs and Uddin, 2013; Turjman, 2016) and in its development from infancy on the basis of contingencies between sensory and motor information (Filippetti et al., 2014; Gallese and Sinigaglia, 2010). Adler et al. (Adler et al., 2016) investigated the mirroring mechanism using somatosensory event-related potentials (SEPs) for self-related information (synchronous visual-tactile stimulation on one's own face) and for other-related information (synchronous visual-tactile stimulation on someone else's face). For people with very infrequent symptoms of depersonalisation (low CDS scores), mirroring effects for self-related information were observed at early stages of somatosensory processing (P45 component, ~45ms post-stimulus), while for other-related information they occurred at later stages (N80 component, ~80ms post-stimulus) (both findings associated with a large effect size). For persons with high levels of depersonalisation (high CDS scores), however, the authors did not observe self-related mirroring effect at P45, but found other-related mirroring at N80 (finding associated with a large effect size). At later cognitive stages (P200 component, ~200ms post-stimulus), mirroring effects for self- and other-related information differed for individuals with low CDS scores but not for those with high CDS scores (findings associated with a large and medium effect size, respectively). In accordance with (Ketay et al., 2014), Adler et al. suggested the lack of early (implicit) self-related information processing as a potential biomarker to explain disembodiment feelings in DPD. A lack of self-other differentiation at later cognitive stages (P200) may also contribute to this aspect of the DPD phenomenology.

In order to explain disembodiment in DPD, it is important to understand how the sense of bodily self-attribution forms in humans. The “rubber hand illusion” (Botvinick and Cohen, 1998;

Ehrsson et al., 2004; Tsakiris and Haggard, 2005) and its full-body virtual reality equivalent (Lenggenhager et al., 2007; Nakul et al., 2020; Pomés and Slater, 2013) have been instrumental in showing that the integration of multisensory information from our environment, specifically our bodily stimuli, forms the sense of bodily self-consciousness (Blanke et al., 2015; Ehrsson et al., 2005; Noel et al., 2018; Tsakiris and Haggard, 2005) (for review see (Tsakiris, 2010)). In the illusion, a participant's real hand receives tactile stimulation synchronously with a rubber hand, while they only see the rubber hand being touched. Synchronous (but not asynchronous) visuotactile stimuli result in changes in body ownership (the feeling that the rubber hand is part of one's own body) and self-location (the felt location of one's own hand shifts toward the rubber hand).

The multimodal integration of exteroceptive signals such as the tactile and visual stimuli employed in the rubber hand illusion and in Adler et al.'s study are thought to occur in several temporal and parietal lobe areas (Calvert and Thesen, 2004), especially the insula (Bushara et al., 2001; Bushara et al., 2003), and disruptions in this process have been proposed as a possible explanation for feelings of disembodiment in DPD (Farmer et al., 2019; O'Sullivan et al., 2018). Indeed, participants with frequent symptoms of depersonalisation (high CDS scores) appear to be more susceptible to feelings of illusory ownership of a rubber hand than those with infrequent symptoms (Kanayama et al. 2009). It is worth noting that this association between biases in multisensory information processing and anomalous bodily experiences is not limited to the clinical population. For instance, Braithwaite et al. (Braithwaite et al., 2017) exposed the rubber hand to a realistic threat following periods of synchronous vs. asynchronous visual-tactile stimulation, and found no differences in GSRs in a group of persons predisposed to out-of-body experiences while a control group showed elevated threat-related GSRs after synchronous compared to asynchronous visual-tactile stimulation.

In addition to the integration of exteroceptive signals, several studies have confirmed the role of interoceptive signals in bodily self-consciousness (Craig, 2009; Tsakiris et al., 2011) (for a review see (Park and Blanke, 2019a)). Interoception refers to the processing of signals originating from visceral organs and represents the internal state of the body (Khalsa and Lapidus, 2016). The main brain regions responsible for interoception are insula, cingulate cortex, amygdala, and somatosensory cortex (Craig, 2009). Notice that the role of these brain regions have been revealed before in depersonalisation (Lemche et al., 2013), and abnormal activities in those brain areas have been observed in DPD patients (Medford, 2012; Medford et al., 2016; Sierk et al., 2018; Sierra et al., 2014). Since activity of insula is attenuated in DPD (Medford et al., 2016; Phillips et al., 2001), attempts to explain disembodiment in DPD based on deficient interoception were made in 2014 (Michal et al., 2014; Sedeño et al., 2014). Interoceptive functioning was evaluated based on two heartbeat detection tasks (Schandry and Whitehead tasks (Schandry, 1981; Whitehead et al., 1977)) in both studies. However, Michal et al.'s DPD patients performed similar to healthy subjects in heartbeat detection tasks while Sedeño et al.'s DPD patients showed impaired performance. This contradiction may be explained by the inability of interoceptive accuracy tests to reliably measure body awareness (Ferentzi et al., 2018). In fact, Michal et al. (Michal et al., 2014) argued that patients with DPD have specific difficulties with sustained attention to interoceptive signals as they show decreasing accuracy in the heartbeat perception task over time, while healthy observers typically increase their accuracy. This pattern may suggest deficits in sustaining attention to bodily sensations, rather than with bodily sensations per se, in patients with DPD. However, a more robust index is needed for the investigation of interoception and bodily self-consciousness in DPD.

One of the main indices of interoceptive signal processing are Heartbeat-Evoked Potentials (HEPs) (Gray et al., 2007; Park and Blanke, 2019b; Pollatos and Schandry, 2004), which correspond to the processing of cardiac signals in the brain. HEPs are obtained by averaging brain signals time-locked with heartbeats in frontocentral regions of the brain, with the insula as their primary origin (Park et al., 2017). They normally can be observed from 200 to 500ms after the occurrence of the R-peak in the human electrocardiogram (Park and Blanke, 2019b; Park et al., 2014a). Changes in HEPs can represent interoception and the level of attention to internal signals. They also reflect distinct attentional mechanisms for interoceptive and exteroceptive signals, such that interoception (tapping in line with one's own heartbeat) yields larger HEP amplitudes compared with exteroception (tapping in line with a recording of a simulated heartbeat) (García-Cordero et al., 2017). This difference was shown to be greater for people who lack interoceptive awareness, as measured through the heartbeat detection task (Schandry, 1981), and who may thus need more attentional effort to perceive their visceral signals (Montoya et al., 1993). In this regard, and to understand the role of interoceptive signals in bodily self-consciousness, Park et al. (Park et al., 2016) investigated the association between HEPs and bodily self-consciousness. They showed that HEPs correspond to the strength of an induced illusory sense of self in a full-body version of the rubber hand illusion. In addition, the authors confirmed the role of insula, cingulate cortex, and amygdala in interoceptive signal processing and bodily self-consciousness, which had been reported before (Heydrich and Blanke, 2013; Ronchi et al., 2015).

Schulz et al. (Schulz et al., 2015) studied the patterns of HEPs in people with DPD and compared them with healthy subjects through a heartbeat perception task. The objective was finding an association between feeling of disembodiment in DPD and the cortical representation of HEPs as an interoceptive signal. Healthy participants showed differential HEP amplitudes after 500ms from the onset of HEP between rest and heartbeat perception task, but such a difference was not found in DPD patients (finding associated with a relatively large effect size). A later study based on the analysis of cardiac responses to startle stimuli demonstrated that the altered pattern of visceral-afferent signals in DPD is not limited to the cortical level, as similar effects were observed in the brainstem (Schulz et al., 2016). It is worth noting that according to their earlier report (Michal et al., 2014), DPD patients' performance in the heartbeat perception task decreased over time. Patients with DPD showed higher initial performance than healthy controls, but showed a decrease over time, while healthy individuals showed an increase over time, resulting in a lack of difference in performance between the two groups overall.

Since it is likely that both interoception and exteroception are involved in the formation of self-consciousness, researchers have attempted to define the role of their integration in bodily self-consciousness (Aspell et al., 2013; Suzuki et al., 2013). For example, Sel et al. (Sel et al., 2017) investigated whether the integration of visual and cardiac information, as an interoceptive and exteroceptive signal respectively, can modify self-face recognition and neural responses to heartbeats. For this purpose, a modified version of the enfacement Illusion (Tsakiris, 2008) as a multisensory integration method was used. The participants saw an unfamiliar face being morphed with their own face and also integrated with a pulsing shade synchronous or asynchronous with their heartbeats. The results showed changes in the HEP amplitude between 195 and 289ms after the R-peak at centroparietal sites of the right hemisphere, with reduced HEP amplitudes during synchronous cardio-visual stimulation compared with asynchronous stimulation. The authors explained the reduced HEP amplitude as a result of a conflict created by the external representation of private information (heartbeats) by an external agent (someone else's face). Based on their argument, the brain resolves this conflict by attenuating the prominence of interoceptive sensations in the formation of self-awareness, which manifests as reduced HEP amplitudes and therefore leads to increased perceived similarity between self and other. Similar results were later obtained by Heydrich et al. (Heydrich et al., 2018) for a cardio-visual stimulation on a full body. Instead of HEPs, they used SEPs as a possible index for changes in self-consciousness. Self-identification with the virtual body modulated SEP component P45, which had previously been shown to reflect reduced implicit self-related information processing in DPD (Adler et al., 2016).

In addition to impairments in the integration of exteroceptive and interoceptive body-related signals, feelings of disembodiment in DPD may be exacerbated by deficits in attentional processes (Adler et al., 2014; Guralnik et al., 2007), and this should be taken into account in future studies of DPD (Gerrans, 2019).

## Emotional numbing (de-affectualisation)

A second core symptom of DPD is emotional numbness (Blevins et al., 2013; Sierra et al., 2002b). Self-reports from patients have asserted the lack of emotional responsivity to external stimuli in DPD (Simeon, 2004), and fMRI studies have shown reduced activity in emotion processing regions such as amygdala, hippocampus, temporal gyrus and anterior insula (Medford et al., 2006). Quaedflieg et al. (Quaedflieg et al., 2013) examined the hypothesis of whether the emotion-induced blindness effect differs in individuals with high versus low levels of depersonalisation determined on the basis of scores on the CDS. Emotion-induced blindness refers to a phenomenon in which one emotionally striking stimulus draws the attention of an individual to such an extent that it reduces the processing capability for further subsequent signals (Most et al., 2005). Due to the lack of patients' emotional responsivity (Simeon, 2004), an inverse relationship between the level of depersonalisation and emotion-induced blindness was expected. The Emotional Scenes Task (Most et al., 2005) was used to present emotional distractors 200ms before the target in order to reduce the ability for correct target detection. As expected, individuals with high levels of depersonalisation (high CDS scores) showed slightly, but not significantly, less emotion-induced blindness than those with low levels of depersonalisation (low CDS scores). The authors also examined visual ERPs during the above task for the two groups. They found a meaningful positive correlation between the magnitude of emotion-induced blindness and the ERP difference wave of emotional versus neutral distractors in a 200–300ms time window at central and frontal electrodes. Interestingly, the ERP difference wave at frontal sites was significantly smaller for high CDS compared to low CDS individuals in the 200–300ms time window (finding associated with a relatively large effect size). Therefore, the authors explained that the impact of an emotional distractor on subsequent processing is less in people with high level of depersonalisation and is found at relatively early stages (200-300ms) of information processing. They also showed that the lack of impact from emotional distractors on these ERP components was associated with the derealisation factor of the CDS rather than with levels of anxiety. In fact, depression and anxiety were related to ERP difference waves in the 600–700ms time window at frontal electrodes. Their findings confirm that DPD is a distinct psychological phenomenon from anxiety or depression (Simeon et al., 2001a; Stanton et al., 2001).

The attenuation of the emotional response in DPD is thought to be caused by decreased activities in emotional cortical (insula) and limbic (hypothalamus, amygdala) regions as well as increased activities of the prefrontal cortex (Lemche et al., 2008; Lemche et al., 2007; Medford et al., 2016; Sierra and David, 2011; Sierra et al., 2002b). The activation of posterior dorsal prefrontal cortex, specifically, is thought to represent true inhibition of the intensity of an emotional stimulus (Konishi et al., 2005; Phan et al., 2005) through the functional coupling between prefrontal and limbic regions (Dolcos et al., 2006). In sum, it is hypothesised that emotional numbing in DPD is due to early overactivation of these prefrontal regions, which triggers the reduction of activity in limbic areas. The activation of this early defensive mechanism could be due to the higher sensitivity of the brain to perceive an external emotional stimulus as a threat (Sierra and Berrios, 1998).

This prefrontal-to-limbic suppressive mechanism can be further investigated using electrophysiological signals. To do so, we suggest that researchers might consider EEG and ERP markers associated with selective cortical inhibition of affective (e.g. aversive) processing, such as alpha wave activity (Uusberg et al., 2013), frontal alpha asymmetries (Palmiero and Piccardi, 2017), delta–beta coherence (Putman, 2011), theta/beta ratio (Putman et al., 2010), and relatively early (200-300ms) posterior negativities in visual ERPs, which are thought to be a consequence of emotional stimulus appraisal in the amygdala (Luu and Tucker, 2003; Schupp et al., 2003), in addition to the frontal and central emotional difference waves in the same time range (200-300ms) that (Quaedflieg et al., 2013) showed to be attenuated in DPD.

## Anomalous subjective recall (de-ideation)

Although functionally intact, memories in DPD can be subject to fragmentation, where patients have difficulty forming sequential and coherent narratives of events (Giesbrecht et al., 2010). DPD patients might also complain of their memories being “colourless” (Sierra and David, 2011, p.4). Although patients can recall autobiographical memories, they describe them as if they did not personally experience them, and as if they were an outside observer of the incident. Sierra and David (Sierra and David, 2011) argued that autobiographical memory recall entails two aspects, including retrieval of the incident and retrieval of the particular feelings during that incident. Although the former aspect is intact in DPD patients, the absence of the latter (Medford et al., 2006) results in actual memories becoming colourless and like a dream. Similarly, a study of visual imagery and perception by Lambert et al. (Lambert et al., 2001b), compared patients with DPD and a group of healthy individuals. DPD patients performed as well as the control group in visual perception tests (Warrington and James, 1991) but showed weakened ability in the imagination of visual information. Since the recall of autobiographical events and the projection of one’s self into an imaginary future are similarly constructive processes (Schacter, 2012), with largely overlapping neural underpinnings in limbic (hippocampus) and medial prefrontal, medial parietal and temporal cortical regions (Martin et al., 2011; Spreng and Grady, 2010), it is conceivable that both may feel similarly self-detached in DPD.

However, and in addition to the questioning of de-ideation as an independent concept in the factor structure of the CDS (Blevins et al., 2013), subjective recall and imagination in DPD have not been adequately studied. In fact, no article was found in the literature on the investigation of de-ideation using electrophysiological signals. However, Papageorgiou et al. (Papageorgiou et al., 2002) observed an altered pattern of P300 during a working memory task in DPD patients. This encoding-related electrophysiological signature could therefore be a potential biomarker to investigate de-ideation in DPD (Amin et al., 2015). Moreover, observation of N200 and frontally distributed N400 (FN400) components of ERP in DPD patients during a memory recall task might also help to discover the underlying nature of de-ideation. A study by Proverbio et al. (Proverbio et al., 2019) showed no distinction in N200 and FN400 components between the retrieval of an old memory and the retrieval of more recent but emotionally salient memory, and both scenarios evoked smaller components in comparison with a recent neutral memory. Since these ERP components index familiarity of a stimulus (Curran, 1999; Curran, 2000), the authors argued that both time and emotional valence have effects on memory recall. Enhanced N200 and FN400 components in DPD patients could therefore mark patients' subjective unfamiliarity with their memory. Finally, an additional biomarker of altered self-related memory and imagination in DPD may be occipital alpha wave activity. Resting-state occipital alpha is an index of visual cortical excitability (Romei et al., 2008), which has recently been associated with individual differences in the strength of visual imagery (Keogh et al., 2020).

## Alienation from surroundings (derealisation)

DPD refers to a chronic condition, and entails, as some researchers proposed (Sierra et al., 2005; Simeon et al., 2008), four distinct symptoms, one of which is derealisation. However, in the case of transient depersonalisation, derealisation is a distinct phenomenon (Dewe et al., 2018), characterised by detachment from surroundings rather than from bodily self, with possible distinct neurobiological mechanisms (Sierra et al., 2002a). For instance, in a recent study by Heydrich et al. (Heydrich et al., 2019), the authors investigated the brain mechanism of depersonalisation- and derealisation-like symptoms in patients with epilepsy. Patients were divided into three groups of those with only depersonalisation-like symptoms, those with only derealisation-like symptoms, and a control group consisting of patients with temporal lobe epilepsy who had experienced Déjà vu or experiential hallucinations. The results from multimodal neuroimaging study revealed that the majority of patients in the first group suffered from frontal lobe epilepsy while the second group mostly suffered from temporal lobe epilepsy. The epileptogenic zone in the group of patients with depersonalisation-like symptoms extended from the mediodorsal premotor cortex to the medial prefrontal cortex. Heydrich et al.'s results thus showed not only that depersonalisation and derealisation are two distinct transient phenomena, but also that they are associated with two different sources of impairments.

No study was found in the DPD literature explicitly targeting derealisation symptoms using electrophysiological methods. We propose that ERP signatures of familiarity and of attentional engagement could serve as potential biomarkers of derealisation. Since derealisation is characterised by a sense of unfamiliarity with one's surroundings, including with spaces and objects that are intimately known, N200 and FN400, which we previously highlighted as familiarity indexes (Curran, 1999; Curran, 2000), may differ less in DPD patients than in controls during exposure to familiar versus unfamiliar scenery. Investigating the allocation of attention within such scenes may also help to delineate the underlying nature of derealisation, as unfamiliar contexts are likely to present a greater source of distraction than familiar contexts (Merriman et al., 2016; Mruczek and Sheinberg, 2007; Park et al., 2014b). P300 has been shown to be associated with cortical engagement in attentional tasks (Polich and Kok, 1995) and may serve as a biomarker for feelings of alienation from one's surroundings. Indeed, abnormal P300 patterns have already been observed in DPD patients in other tasks (Papageorgiou et al., 2002; Wise et al., 2009). We further propose EEG/ERP indices of spatial cognition within egocentric and allocation reference frames as potential measures of derealisation symptoms. For example, egocentric (vs. allocentric) encoding of object locations in space has been associated with larger N1 amplitudes and longer N2 latencies at left and bilateral inferior parietal sites, respectively (Lithfous et al., 2014), possibly resulting from differential spatial discrimination and frame-dependent localisation processes at these stages (Hopf et al., 2002; Luck and Hillyard, 1994; Vogel and Luck, 2000). Gramann and colleagues (Gramann et al., 2010) found that the use of an egocentric (vs. allocentric) reference frame during spatial navigation was associated with greater alpha suppression in or near right primary visual cortex (vs. in occipito-temporal, bilateral inferior parietal, and retrosplenial cortical areas). As a result of incoherent spatial referencing between body and environment, such EEG/ERP signatures may be expected to be altered in persons experiencing derealisation symptoms.

Another way to investigate derealisation (and other DPD) symptoms may be through the probing of the vestibular system. The vestibular system includes sensory organs located in the inner ear, which send information about the head's position, spatial orientation, and motion to the brain (Khan and Chang, 2013), where vestibular signals are processed in distributed regions from the temporo-parietal cortex to the prefrontal cortex (Ventre-Dominey, 2014). The vestibular system plays a crucial role, not only in the sense of balance, motor coordination and spatial orientation, but also more broadly for egocentric self-awareness (Lenggenhager et al., 2008; Lenggenhager et al., 2015) by providing a gravitational reference to other bodily signals (Ferrè et al., 2014). When signalling pathways are disturbed in peripheral vestibular disease, the brain fails to generate a coherent spatial representation of the body with respect to the external world (Renaud, 2015). An incoherent spatial frame of reference may also result in feelings of detachment from the world (derealisation). Several studies in the literature indeed report a higher tendency for DPD symptoms among patients with peripheral vestibular disease than among healthy individuals. Sang et al. (Sang et al., 2006) first showed this in a sample group of 171 subjects (121 healthy, 50 patients), and later (Jáuregui-Renaud et al., 2008) reported that levels of depersonalisation were negatively correlated with patients' ability to estimate spatial orientation in an environment. Further substantiating the potential involvement of vestibular signals in DPD symptoms, Tschan et al. (Tschan et al., 2013) found that detachment from memory, derealisation and disembodiment in the general population were the three most substantial DPD symptoms associated with feelings of vertigo and dizziness, the most frequent symptoms seen in vestibular patients (Sang et al., 2006).

Future studies of DPD, specifically those interested in derealisation, may thus consider activating the vestibular system through passive full-body motion or through direct artificial stimulation of the (otolith) vestibular system (Ertl and Boegle, 2019), and probing EEG / ERP markers pertaining to cortical vestibular processing (vestibular evoked potentials, EEG power and EEG microstates). For instance, several vestibular evoked potentials can be measured over posterior, frontal and central sites within 500ms of passive motion, acoustic or galvanic stimulation (Ertl et al., 2017; Kammermeier et al., 2017; Kammermeier et al., 2015). Evoked potentials can reflect different motion parameters and have been source-localised to the cingulate sulcus visual area and the opercular-insular region (Ertl et al., 2017). These studies have also identified evoked beta- and mu-band responses in central electrodes. EEG studies have further shown that motion-induced vestibular stimulation caused bilateral temporal-parietal suppression of alpha oscillatory activity (Gale et al., 2016; Gutteling and Medendorp, 2016), which was found to be attenuated in patients with vestibular loss (Gale et al., 2016). Cortical EEG/ERP signatures like these may thus be useful for studying the vestibular system in DPD, and egocentric self-awareness in general.

## Other symptoms of DPD

Standard neuropsychological test have detected broad perceptual and attentional alterations in the pathophysiology of depersonalisation/derealisation (Guralnik et al., 2007; Guralnik et al., 2000). Already Hollander et al. (Hollander et al., 1992) suggested that depersonalisation is marked by dysfunctions in the emotional modulation and integration of perceptual information, and they reported increased absolute values of visual and auditory N200 components of event-related potentials (ERPs) over the left temporal cortex in a DPD patient, in addition to abnormal theta oscillatory activity.

An electrophysiological alteration in early attentional functioning in DPD was later verified in (Schabinger et al., 2018). Schabinger et al. used a spatial cueing paradigm (Posner and Cohen, 1984) to investigate the selective attentional mechanisms (Hillyard and Anllo-Vento, 1998; Posner et al., 1980) in DPD and psychosomatic control patients, who were matched for depression and anxiety. Each trial in this paradigm consisted of a target (ellipse) or non-target (circle) visual stimulus in either the left or right area of the screen proceeded by a central spatial cue (arrow) indicating correctly, incorrectly, or neutrally the location of the upcoming stimulus. Visual ERPs in response to cued stimuli were investigated for the two groups. Schabinger et al. found diminished suppression of stimuli at to-be-ignored locations at the early sensory P1 component in DPD patients compared to control patiends (findings associated with a large effect size) while attentional effects at sensory N1 (enhancement of stimuli at to-be-attended locations) and later cognitive components (including P300) were similar across patient groups. It was proposed that the lack of early-stage suppression of irrelevant sensory inputs might be responsible for the distractibility reported by DPD patients. Schabinger et al suggested visual ERP component P1 as a potential biomarker for deficient attentional functioning in chronic DPD.

Papageorgiou et al. examined the potential alteration in the P300 ERP component in transient depersonalisation (Papageorgiou et al., 2002). In a working memory test, lower P300 amplitudes (but no changes in latency) at central posterior brain regions were observed in individuals with transient depersonalisation experiences compared with a control group (finding associated with a large effect size). Since high P300 amplitudes are typically evoked by conditions which demand more attention (Holdstock, 1995; Polich, 1998), the findings of the above study appear to confirm attention problems in depersonalisation states. However, a later study (Wise et al., 2009) reported a related but inverse finding regarding amplitude and latency changes in P300 in depersonalisation state. Analysis of ERP components during an auditory oddball task in patients with panic disorder revealed that patients who had experienced depersonalisation symptoms showed reduced P300 latency (but no changes in amplitude) to a striking target sound compared with healthy individuals (finding associated with a medium effect size). Reduced P300 latency indicates accelerated information processing and stimulus classification (Picton, 1992; Sur and Sinha, 2009). No such reduction in P300 latency was observed in the comparison between panic disorder patients without depersonalisation and the control group. In addition to the unchanged attentional effects at P300 in a visual-spatial task comparing DPD with psychosomatic control patients (Schabinger et al., 2018), longer (rather than reduced) P300 latencies were reported in panic disorder patients compared to healthy individuals (Turan et al., 2002). In sum, although P300 may be a valuable electrophysiological biomarker for attentional deficits in depersonalisation/derealisation, more studies with carefully designed tasks are needed to examine its precise expression in each task and for each group of patients. Additional ERP biomarkers for abnormalities in perceptual-attentional systems in DPD may be found in P1 (Schabinger et al., 2018) and N200 (Hollander et al., 1992) components, and these also require substantial replication through new studies of transient and chronic DPD.

## Summary and Conclusions

Depersonalisation/derealisation disorder (DPD) can profoundly affect the quality of life for patients and interfere with their social relationships and daily activities. It usually takes several years to be correctly diagnosed (Hunter et al., 2017), and the symptoms of the disorder can be intolerable until then. The present paper provided a systematic review of the studies targeting transient and chronic symptoms of depersonalisation using electrophysiological neuroimaging techniques. The aim was to describe what is presently known about the neurophysiological correlates of DPD symptoms and to make recommendations for further study to improve the diagnostic potential of this neuroimaging tool.

Before we summarise the EEG/ERP indices of DPD, it is worth noting the sparse use of electrophysiology to delineate the neurophysiological correlates of DPD symptoms. Only ten studies satisfied our criteria for inclusion in this review, two of which were single-case studies. Yet, EEG/ERP methodologies are powerful techniques for eliciting human perception, cognition and action independently of participants' cultural background or education levels. Unlike fMRI, EEG/ERPs are direct measures of neural activity. Inexpensive, easy to implement, and well tolerated by patients, their diagnostic potential has been successfully studied in other disorders, including schizophrenia (Luck et al., 2011), which has a similar prevalence as DPD. We would therefore urge researchers interested in depersonalisation/derealisation to invest more resources into these techniques to develop their diagnostic potential for DPD beyond that which is presently known.

Several studies have shown abnormal EEG activities in theta band in DPD patients, and the severity of symptoms was found to be associated with increased theta activity. Higher theta wave synchronisation has been observed for emotional compared to neutral stimuli, and this synchronisation occurs earlier (around 200ms) when emotional stimuli are processed implicitly but later (around 300ms) during the explicit recognition of facial emotions (Knyazev et al., 2009). Therefore, increased theta activity in DPD patients might be associated with their greater effort or involvement in processing emotional information (specifically unpleasant emotional information (Quaedflieg et al., 2013; Sierra et al., 2002b; Sierra et al., 2006)). Another study (Kirmizi-Alsan et al., 2006) showed that the theta wave represents the suppression of nontarget stimuli in the go/nogo task, which requires both selective inhibition and arousal. Therefore, theta activity might also represent the greater effort in the selective suppression of processing, which has been observed to be deficient in DPD patients (Schabinger et al., 2018). Theta further plays a role in memory maintenance in that higher theta activity is associated with the need for greater working memory capacity (Jensen and Tesche, 2002; Raghavachari et al., 2001). Thus, the theta band of the EEG power spectrum is likely to serve as a potential electrophysiological biomarker to study DPD symptoms related to emotion, attention/inhibition and working memory. Future studies of DPD should therefore focus on this oscillatory signature and investigate (a) the temporal dynamics of event-related synchronisation in response to emotional stimuli, and (b) oscillatory power during transient episodes of depersonalisation and in patients with chronic DPD, for example. Note that it is important for future studies to consider a relatively prolonged time window in order to analyse low-frequency components of EEG such as theta (practically, the time window needs to contain at least three cycles of the target frequency), which may affect stimulus design.

Analysis of cortical event-related and heartbeat-evoked potentials (ERPs and HEPs) can also reveal valuable information regarding the underlying nature of DPD symptoms, as well as of the sensorimotor integrative processes contributing to bodily self-consciousness in general. HEPs and some ERP components have been introduced as potential valuable indices to investigate DPD symptoms related to disembodiment. Somatosensory P45 reflects processing in the primary somatosensory cortex and is known to be involved in attributing body ownership (Adler et al., 2016; Otsuru et al., 2014; Rigato et al., 2019). Cortical HEPs are thought to primarily derive from insula activity, which is attenuated in DPD (Medford et al., 2016; Phillips et al., 2001). A lack of P45 modulation during visual-tactile stimulation related to the self (Adler et al., 2016), and a lack of HEP modulation during focused attention to one's own heartbeat (Schulz et al., 2015), may both serve as an electrophysiological biomarkers for feelings of disembodiment (desomatisation) in transient and chronic DPD. Both somatosensory cortex and insula are part of the networks responsible for interoception (Craig, 2009), and the integration of interoceptive, exteroceptive, and interoceptive with exteroceptive sensory signals have been recently proposed as critical for generating the moment-to-moment feeling of self-consciousness. When integrative processes like these are transiently or chronically dysfunctional, feelings of disembodiment may ensue, and may be further exacerbated by abnormal attentional processes (Gerrans, 2019). Our review has highlighted the potential of somatosensory P45 and of HEPs for measuring these processes in health and disease. It is worth mentioning that there might be a link between somatosensory P45 and the cortical HEP in the investigation of DPD symptoms. A recent study (Adler and Gillmeister, 2019) showed that the activation of somatosensory P45 component in response to personal visual-tactile stimulation (touch on subject's own hand) is stronger in people with a higher level of interoceptive awareness. If interoceptive awareness modules HEPs (Petzschner et al., 2019), measuring both P45 and HEPs in the same study may be able to test the relationship between interoceptive and exteroceptive awareness (Al et al., 2019). In addition, researchers may consider the inclusion of electrophysiological markers for proprioceptive and vestibular signal processing to investigate the bodily self in health and disease. Vestibular processes are increasingly recognised as critically involved in feelings of body ownership and egocentric perception and vestibular disturbances bear a strong link with several cardinal DPD symptoms (Lenggenhager et al., 2015; Tschan et al., 2013).

Another potential set of ERP biomarkers for DPD symptoms occur around 200-300ms (Quaedflieg et al., 2013), including N200 (Hollander et al., 1992), P200 (Adler et al., 2016) and P300 (Papageorgiou et al., 2002; Wise et al., 2009). In this article, we have referred to enhanced temporal N200 as an aspect of alterations in the perceptual-attentional system (Hollander et al., 1992), to lack of self-other differentiation in somatosensory resonance at frontocentral P200 as an aspect of disembodiment (Adler et al., 2016), to lack of frontocentral 200-300ms differences (Quaedflieg et al., 2013) as an aspect of emotional numbing, and to reduced amplitude/latency at centro-parietal P300 (Papageorgiou et al., 2002; Wise et al., 2009) as an aspect of working memory/attentional dysfunction. We have also suggested the potential of additional markers in this time range, such as the N200 and frontal N400 related to stimulus familiarity. Partly due to inconsistencies in the precise expression of P300 abnormalities in DPD, further studies are urgently needed to systematically confirm P300 as a potential electrophysiological biomarker for diagnosing and investigating DPD symptomatology related to attentional and working memory dysfunction, as well as to symptoms of de-ideation and derealisation. Caution should be applied when regarding the association between P300 and DPD symptoms because P300 changes also occur with a number of other pathologies and physiological states (e.g. (Duncan et al., 2009; Gangadhar et al., 1993; Mathalon et al., 2000)) and are thus not necessarily selective for depersonalisation.

Reduced brain activities have been observed in DPD patients in sensory information processing units (Medford et al., 2016) as well as regions responsible for the processing of visceral signals (Lemche et al., 2013). Besides, the impairment is found mainly in the early stages of information processing (Adler et al., 2016; Quaedflieg et al., 2013; Schabinger et al., 2018). This impairment in the implicit processing of multimodal interoceptive and exteroceptive signals could be a result of long-time severe stress or anxiety (Shilony and Grossman, 1993), which may damage the sensory processing units and reduce their processing capacity (Sierra and Berrios, 1998). For instance, one dominant theory to explain emotional numbing in DPD (Sierra and Berrios, 1998) defines a threshold for the level of anxiety (or any unpleasant salient stimuli) after which the emotional processing units (including the anterior insula and amygdala) discontinue translating emotions into perceived feelings, and DPD is associated with abnormalities in this threshold or in how quickly it is reached (Jay et al., 2014). The higher GSR baseline and the earlier peak in GSR response of DPD patients to emotional stimuli in (Giesbrecht et al., 2010) represent a faster saturation of emotional capacity (Sierra et al., 2002b). Further, reduced capacity in sensory processing units may also be the cause of concentration problems in DPD, in a way that there may be a commensurate loss in the capacity to filter relevant from irrelevant signals in sensory information. A related framework within which symptoms of DPD may be explained is based on the loss of the brain's ability to make and update predictions about the internal body state and the outside world (Gerrans, 2019; Seth et al., 2012; Seth and Tsakiris, 2018). With reduced activity in emotional information processing units (such as insula), the brain faces a relative sparsity of information for maintaining precision in predictions about one's self and the outside world. The increasing inconsistencies may cause further suppression of emotional processing and, in turn, give rise to DPD symptoms such as the feeling of disembodiment or derealisation. Conflict-monitoring related electrophysiological markers such as the mismatch negativity (MMN) and the error-related negativity (ERN) are good candidates for investigating this theory further. MMN is a pre-attentive fronto-central response around 100-200ms after omitted or perceptually deviant sensory events in a stimulus sequence, and is generated by a network of temporal-prefrontal cortical regions (Garrido et al., 2009). ERN is generated by a region in the anterior cingulate cortex (ACC) (Dehaene et al., 1994) and emerges around 100ms after the onset of an erroneous motor response, even when the observer is unaware of making an error and even when the error cannot be corrected (Luu and Tucker, 2003). Recent studies have begun to describe both MMN and ERN within a predictive-coding framework as indexes of the brain's monitoring of bodily processes, where MMN may represent the failed prediction of visceral and other sensory inputs and thus the failed suppression of the prediction error (Garrido et al., 2009; Pfeiffer and De Lucia, 2017). ERN may represent the monitoring of errors committed as a consequence of failed interoceptive and exteroceptive predictions, with imbalances in monitoring shown to be related to anxiety pathology (Sueyoshi et al., 2014; Tan et al., 2019; Yoris et al., 2017). Thus, MMN and ERN may be useful for indexing errors in predicting the internal and external state of the body in persons with DPD.

There is not enough evidence to confirm a clear hemispheric lateralisation of DPD symptoms' neurophysiological correlates. Nevertheless, it is likely that such hemispheric biases will eventually emerge with more targeted research. Some studies have already shown abnormalities in left-hemispheric activation of DPD patients (Hollander et al., 1992; Jiménez-Genchi, 2004) and of individuals with more frequent dissociative experiences (Spitzer et al., 2004). Furthermore, based on the dominant role of the right hemisphere in emotional (Gainotti, 2019) and self-related processing (Hu et al., 2016), right-hemispheric biases in the dysregulation of emotional processing and disembodiment in DPD may be expected. In support of this argument, it may also be interesting to note that the right hemisphere may be more involved in the perception and processing of negative emotions, while the left hemisphere deals predominantly with positive sensations (Silberman and Weingartner, 1986), and that impairment in emotional responsivity in DPD is often exclusive to unpleasant and threatening stimuli rather than pleasant ones (Michal et al., 2013; Monde et al., 2013; Sierra et al., 2006). In addition, inhibitory rTMS applied to the right ventrolateral prefrontal cortex can increase arousal capacity (Jay et al., 2014) and rTMS applied to the right temporoparietal junction can reduce symptoms (as measured by CDS total score) in DPD patients (Mantovani et al., 2011). Increased activation in the angular gyrus of the right parietal lobe has also been found to be correlated with the level of depersonalisation (Simeon et al., 2000). However, the DPD literature currently faces a lack of direct investigations of hemispheric differences related to DPD symptoms; more studies on this are urgently needed.

A schematic of the notable biological signals and associated electrophysiological biomarkers introduced in this review as potential diagnostic tools are depicted in **Error! Reference source not found.**. Future studies in this area should consider both indexes of interoceptive (such as HEPs) and exteroceptive signals (such as visual or somatosensory ERPs), vestibular signals, and especially their interaction (including monitoring ERP markers like MMN, ERN). For a comprehensive picture of DPD, it is useful to jointly investigate both peripheral (autonomic) and central (cerebral) bodily responses, and both early (perceptual) and later (cognitive) stages of central information processing. Additionally, there is a lack of research in the analysis of EEG power spectra and their relative ratios and coherence. As this review has uncovered that these may yield promising biomarkers for all cardinal DPD symptoms, further studies need to consider different EEG waveforms and their roles in the formation of these symptoms. For such analyses, the use of time-frequency and phase-based signal processing (Mowlaee et al., 2016) can be very insightful, since studies have suggested phase-alignment as a fundamental phenomenon underlying the generation of ERPs and HEPs rather than evoked potentials (Burgess, 2012; Park et al., 2018).

Further, the association between the vestibular system and the DPD symptoms can help the design of experimental paradigms in future studies (Lopez et al., 2018). In this regard, a study by Sang et al. (Sang et al., 2006) proposes caloric vestibular stimulation as an effective way to provoke depersonalisation/derealisation-like symptoms in the non-clinical population. The induced symptoms were similar to those experienced by patients with vestibular disease. However, it should be carefully considered that there is also a close link between the vestibular system dysfunction and out-of-body experiences (Blanke et al., 2004; Lopez and Elziere, 2018; Lopez et al., 2008). For example, Lopez et al. (Lopez and Elziere, 2018) reported a higher tendency for out of body experiences in patients with dizziness, and the relationship was more significant in patients with peripheral vestibular disorder. Since out-of-body experiences are not that common in depersonalisation/derealisation (Sierra and David, 2011), using methods to dysregulate the vestibular system (such as caloric vestibular stimulation) to induce depersonalisation/derealisation symptoms should be done carefully to prevent misinterpretation.

A correct diagnosis of DPD is an urgent matter in the area of psychological disorders, and there is a need for finding diagnostic markers highly specific to DPD in order to distinguish it from other alternative diagnoses. Understanding the potential of electrophysiological tools for the identification of DPD symptoms can help to diagnose DPD quickly and effectively, and we therefore appeal to researchers interested in the phenomena of self-awareness in health and disease to consider using these tools more frequently.

# Declaration of Conflicting Interests

The author(s) declared no potential conflicts of interest with respect to the research, authorship, and/or publication of this article.

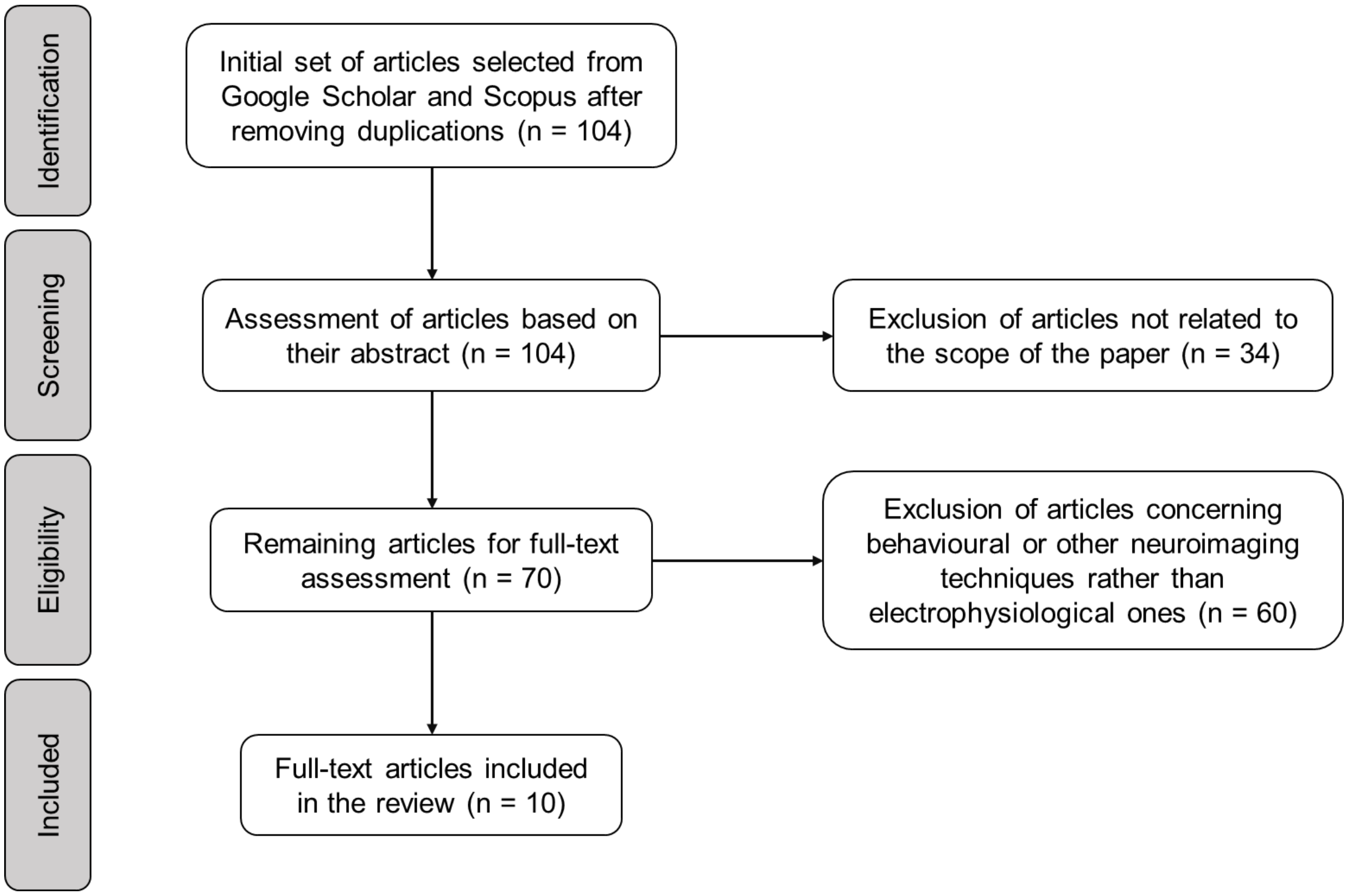

**Figure 1** Flow diagram of the article selection process through the systematic review

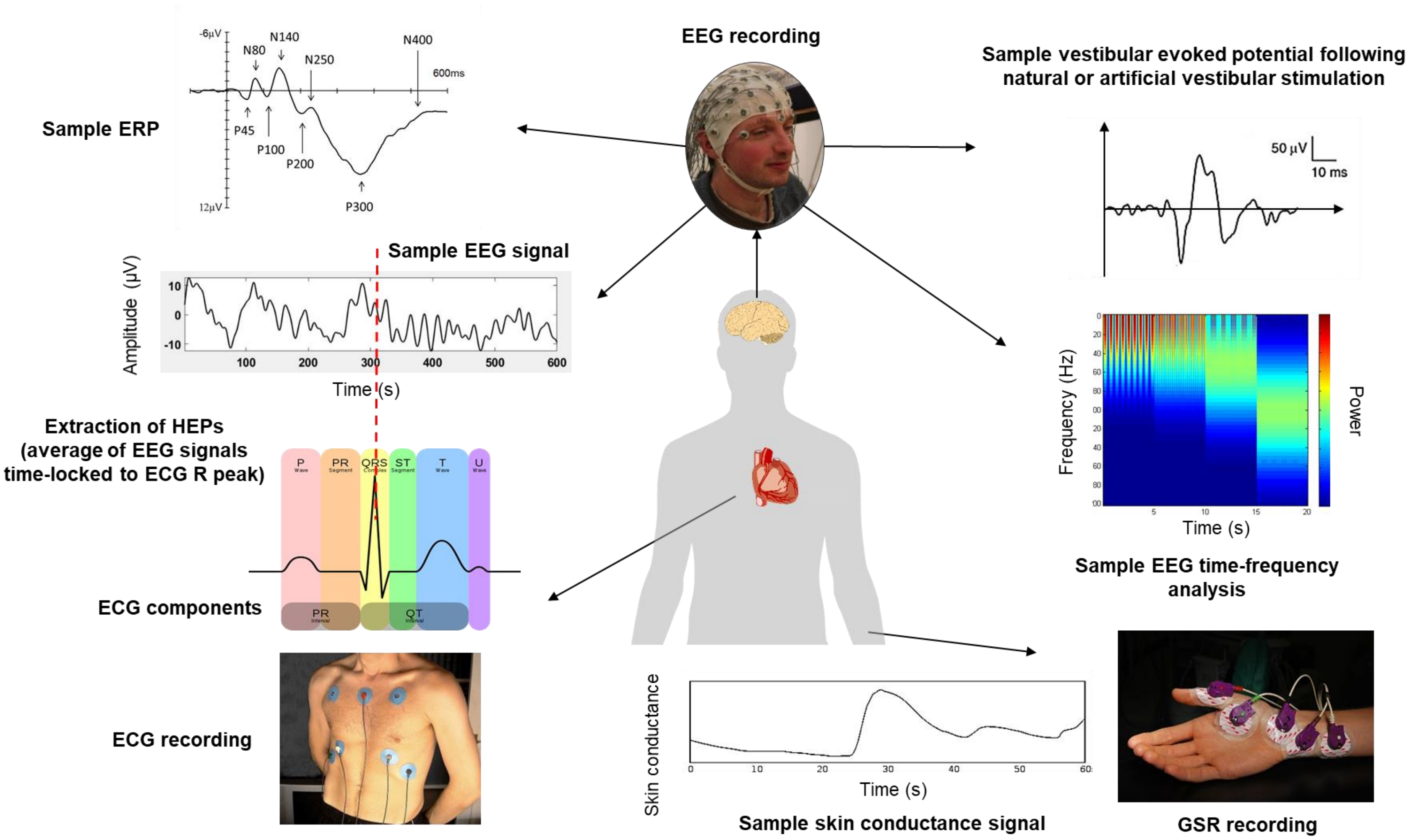

**Figure 2** An overview of the biological signals and relevant electrophysiological biomarkers introduced in this review

**Table 1** Four major symptoms of DPD (adapted from (Medford et al., 2005; Sierra and David, 2011)) and other associated processing differences

| | |
|---|---|
| Disembodiment feelings (desomatisation) | Lack of body ownership or loss of agency |
| Emotional numbing (de-affectualisation) | Attenuation in emotional responsiveness |
| Anomalous subjective recall (de-ideation) | Disassociation between an incident and personal feeling in autobiographical memory retrieval |
| Alienation from surroundings (derealisation) | Detachment of the self from its surroundings |
| Other symptoms and processing differences | Impaired attentional functioning and processing speed or perceptual organisation |

**Table 2** Studies using electrophysiological neuroimaging techniques to characterise transient and chronic depersonalisation symptoms. Note that studies are ordered according to target symptom, and then chronologically. Reported effect sizes (interpretation): *d*: Cohen's d effect size (small (0.2), medium (0.5), and large (0.8)). $\eta_p^2$: partial eta squared effect size (small (0.01), medium (0.06), and large (0.14)). OR: odds ratio effect size (small (1.5), medium (2.5), and large (4.3)).

| Study (number of participants) | Task | Electrophysiological biomarker | Target symptom (transiency) | Findings |
|---|---|---|---|---|
| (Hollander et al., 1992) (n = 1) | None specified (a case report) | EEG power spectrum (alpha, theta); ERPs (visual and auditory; cognitive N200) | None specified (chronic) | Increased frontal alpha and temporal theta activity and enhanced N200 in visual and auditory ERPs over left brain regions in a DPD patient (case study). |
| (Locatelli et al., 1993) (n = 37) | Odour discrimination task | EEG power spectrum (delta, upper alpha) | None specified (transient) | Increased power in the delta band and a bilateral lack of responsiveness in the upper alpha band during odour discrimination in panic disorder patients with versus without depersonalisation (unable to obtain effect size from available data). |
| (Raimo et al., 1999) (n = 1) | None specified (a case report) | EEG power spectrum (theta) | None specified (transient) | Significantly higher relative power in theta band during induced depersonalisation episodes by absolute alcohol consumption compared with asymptomatic episodes (case study). |
| (Hayashi et al., 2010) (n = 70) | Intermittent Photic Stimulation (IPS) | EEG power spectrum (theta) | None specified (transient) | Abnormal theta activity (repeated slow waves) in 17 of 70 panic disorder patients, which was associated with three panic disorder symptoms (depersonalisation, nausea, and paresthesias) (OR = 13.92 for association with depersonalization/derealization). |
| (Schulz et al., 2015) (n = 47) | Heartbeat perception task | HEPs | Disembodiment (chronic) | Reduced differential HEP amplitudes after 500ms between rest and heartbeat tracking in DPD patients (n= 23) compared with healthy controls (n = 24) ($\eta_p^2$ = 0.10 for the two-way interaction between group and experimental task). |
| (Adler et al., 2016) (n = 27) | Somatosensory resonance paradigm | ERPs (somatosensory; perceptual P45, cognitive P200) | Disembodiment (transient) | Reduced somatosensory resonance for self-related information at P45 (no changes for other-related information at N80) in central-parietal electrodes in individuals with high (n = 14) versus low (n = 13) CDS scores (P45: $\eta_p^2$ = 0.46 for self-face touch in low CDS group only; N80: $\eta_p^2$ = 0.29 for other-face touch in all participants). Reduced self-other differences in resonance at cognitive P200 in fronto-central electrodes in individuals with high versus low CDS scores ($\eta_p^2$ = 0.06 versus $\eta_p^2$ = 0.41 for interaction between observed touch and observed face in each group). |
| (Quaedflieg et al., 2013) (n = 30) | Emotion-induced blindness (EIB) | ERPs (visual; cognitive 200-300ms) | Emotional numbing (transient) | Smaller difference in ERP amplitudes in a 200–300ms time window between emotional and neutral distractors in frontal electrodes in individuals with high (n = 15) versus low (n = 15) CDS scores (*d* = 0.71 for differential activation at Fz in 200–300ms time window between the two groups). |
| (Papageorgiou et al., 2002) (n = 30) | Working memory test | ERPs (auditory; cognitive P300) | Other (working memory) (transient) | Lower P300 amplitudes (no changes in latency) in response to single sound presenting the start of working memory tests at central posterior brain regions in healthy individuals with transient depersonalisation (n = 15) compared with a control group (n = 15) (*d* = 1.62 for group difference in P300 amplitudes). |
| (Wise et al., 2009) (n = 75) | Auditory oddball task | ERPs (auditory; cognitive P300) | Other (attentional dysfunction) (transient) | Reduced P300 latency (no changes in amplitude) in response to striking (target) tone in panic disorder patients with transient depersonalisation (n = 25) compared with matched healthy controls (n = 50) (*d* = 0.62 for group difference in P300 latency). |
| (Schabinger et al., 2018) (n = 28) | Spatial cueing paradigm | ERPs (visual; perceptual P1) | Other (attentional dysfunction) (chronic) | Reduced attentional suppression of irrelevant sensory inputs at P1 in occipito-parietal electrodes in DPD patients (n = 14) compared with anxiety- and depression-matched psychosomatic patients without DPD (n = 14) ($\eta_p^2$ = 0.66 vs. $\eta_p^2$ = 0.19 for effect of cue validity (valid vs .neutral vs. invalid) in DPD vs. control group). |